\documentstyle[epsf,referee]{l-aa-mpa}
\def\overmag#1.#2 {\hbox{$#1^m$\kern -6.0pt .\kern 3.0pt #2 }}  %
\def\mathrm{\rm}

\begin{document}

\thesaurus{05(08.05.3; 08.12.3; 10.07.2)}

\title{The level of agreement between theoretical and observed
       globular cluster luminosity functions}

\author{S.~Degl'Innocenti\inst{1,2} \and A.~Weiss\inst{1} \and
L.~Leone\inst{3}}

\institute{Max-Planck-Institut f\"ur Astrophysik,
           Karl-Schwarzschild-Str.~1, 85740 Garching,
           Federal Republic of Germany
           \and
           INFN Sezione di Ferrara, I-44100 Ferrara, Italy
           \and 
           Dipartimento di Fisica, Universit\`a di Pisa,
           Piazza Torricelli 2, I-56100 Pisa, Italy
           }

\offprints{A.~Weiss}

\date{Received; accepted}

\maketitle

\begin{abstract}
Luminosity functions from theoretical stellar evolution calculations
are compared with observed ones of several galactic globular clusters
(M30, M92, M68, NGC6397, M4, M80, NGC6352, NGC1851). Contrary to
earlier results of Faulkner \& Stetson (1993) and Bolte (1994) we find
no significant discrepancy that could indicate the neglect of
important physical effects in the models. However, it is confirmed
that the subgiant branch is the most sensitive part and shows the largest
deviations in the 
luminosity function comparison, if parameters are unappropriate. 
We also find that the main sequence is suited less than the Red Giant
Branch for the calibration of theoretical luminosity functions.
While for individual clusters 
different changes in the model assumptions might resolve mismatches,
there is no systematic trend visible. It rather appears that the
quality of the luminosity function in the subgiant part is
insufficient and that  
improved observations of this particular region are necessary for a
better comparison. At the 
present quality of luminosity functions theory is 
in agreement with observations and a postulation of WIMPs acting in
stellar cores does not seem to be justified. Actually, fits using
isothermal core models on the main sequence appear to be worse than
those with standard stellar evolution assumptions.
\end{abstract}

\markboth{Degl'Innocenti et al.: Globular cluster luminosity functions}{}

\newpage 

\thispagestyle{plain}
\mbox{~}\par

\bigskip\bigskip

\centerline{\Large\bf The level of agreement between theoretical and observed}
\centerline{\Large\bf globular cluster luminosity functions}
\medskip
\centerline{\sl Scilla Degl'Innocenti, Achim Weiss, and Luigi Leone}
\bigskip

\section{Introduction}

Globular clusters are compared with the theory of stellar structure
and evolution by use of two functions: the colour-magnitude-diagram
(CMD) displaying the surface properties of the stars and the
luminosity function (LF) measuring the initial mass function (IMF) for
the unevolved lower main sequence and the speed of evolution after the
turn-off. According to Faulkner \& Swenson (1993) the agreement with
the observed LF of the globular clusters M30, M92, M68 and NGC6397 is
sufficiently bad and unresolvable with standard stellar structure
physics that an additional physical mechanism acting during the
main-sequence evolution has to be postulated. In particular, they
found that the observed LFs show an excess of stars on the subgiant
branch, which is interpreted as a prolonged subgiant phase resulting
from a higher central hydrogen abundance at the turn-off and an
initially broader hydrogen burning shell, which is getting very narrow
during the subgiant phase.
The shell, 
developing at the end of the main sequence, can be made broader, if
some part of the stellar core would be isothermal. Faulkner \& Swenson
(1988, 1993) investigated the evolution of low-mass stars on the main
sequence and the subsequent subgiant and giant branch under the
assumption of such an isothermal core extending over the innermost
10\% of the stellar mass and found a better agreement between
theoretical and observed LF. Bolte (1994) confirmed these results for
M30 using an improved $V$-band LF but the same theoretical models
(Bergbusch \& VandenBerg 1992).
As a side-effect of the isothermal cores, globular clusters would be
younger by about 20\% than usually thought (Faulkner \& Swenson 1993).

The origin of the efficient energy transport needed in the stellar
center was suggested to be found in the presence of WIMPs accreted by
the star during its main-sequence phase (Faulkner \& Swenson
1993). Bolte (1994) already mentioned that only a very small region in
the WIMPs' mass--cross-section--phasespace is left over from various
experiments and theoretical expectations. In addition, solar models
without an isothermal core better fit the helioseismological data (Cox,
Guzik \& Raby 1990; Cox \& Raby 1990; Kaplan et al.~1991). A similar
result is reported by Basu \& Thompson (1996) for a solar seismic model
with a reduced central temperature simulating the effect of some
additional energy transport by WIMPs. 
These and additional arguments led Bolte to the
conclusion that WIMPs are unlikely to be a major constituent of halo dark
matter accreted by stars and affecting their core structure.
Since his paper the parameter space for WIMPs -- when assumed that they
are neutralinos -- has constantly been
shrinking (Fornengo 1994; Mignola \& Berezinsky, private
communication); thus his conclusion is more justified 
than ever. However, depending on new ideas the nature of WIMPs
might be different and they possibly might exist and act in stars.

Independent of this, the result of Faulkner \& Stetson (1993) and Bolte
(1994) remains: that there was an apparent mismatch between theoretical and
observed GC luminosity functions, which can be reconciled by an
isothermal core. The reason for the isothermality would remain
unclear.
Since this discrepancy is a severe challenge to stellar structure
theory, we re-investigated the quality of both theoretical and
observational LFs. In Sect.~2 we will discuss the method of obtaining
a theoretical LF and problems and errors associated with the observed
LFs. In the following section, we will perform standard comparisons
with the Faulkner \& Stetson and other clusters and discuss the
results. In Sect.~4 we will investigate variations in the standard
physics and the effect of  isothermal cores. At last, our conclusions
follow in Sect.~5. 

\section{Luminosity functions: observed and calculated}

To compare theoretical luminosity functions with the
observed ones of different globular clusters we need isochrones
for various ages, chemical compositions and initial mass functions
(IMF).
With these quantities we obtain the number of stars in any given interval
of visual magnitude by
$$
\frac{dn}{dM_V} = \frac{dn}{dm} \cdot \frac{dm}{dM_V}.
$$
For the IMF we use the classical power law form: 
$$
\frac{dn}{dm}= m^{-s},
$$
with $m$ being the stellar mass.
To find the best agreement
with observational results  we tried to fit the data of each cluster
with several luminosity functions for different ages, metallicities and
exponents for the power law of the IMF (we recall that the IMF affects
the MS only); the best fit for each cluster will be shown
in the next section. All LFs are normalized to the total number of stars on
the RGB. We have checked that the normalization does not depend on the
brightness range used.
As will be demonstrated in Sect.~3, the RGB-part of the LF is not at
all influenced by any assumption we have tested, including that of an
isothermal main-sequence core. In contrast, the main-sequence part
depends strongly on the IMF and -- though less -- on the helium
content. It is therefore natural to prefer the RGB for the
normalization of the LFs. 
In some cases we also allowed a horizontal (i.e.\ in visual
magnitude) shift in the LF of 0\fm1, which is certainly less than the
uncertainty in the adopted distance modulus.

The helium mass fractions of all  isochrones is $Y=0.23$ or
$Y=0.24$. This small difference does not influence the LFs (Sect. 4.3\
and Ratcliff 1987)
The range of age and metallicity covered by the adopted isochrones are
 10 $\div$ 20 Gyr and $Z=0.0001 \div 0.006$, resp.
We only used solar metal ratios within $Z$. Salaris, Chieffi \&
Straniero (1993) have demonstrated that for metal poor stars only the
total or {\sl global} metallicity is important for evolution,
isochrones and therefore LFs. We thus can savely ignore
$\alpha$-element enrichment, which is well observed in many GCs.

All evolutionary calculations have been made with the Frascati
Raphson Newton Evolutionary Code (FRANEC) whose general features and
physical inputs has already been described in previous papers (see
e.g. Chieffi \& Straniero 1989). For all metallicities except $Z=0.0002$
we adopted the isochrones of Chieffi \& Straniero (1989) and
Straniero and Chieffi (1991) with radiative opacity coefficients from the
Los Alamos opacity library (Huebner et al. 1977; Ross \& Aller 1976 solar
metal ratio), combined with the
Cox \& Tabor (1976) opacities in the low temperature region (below
$10^4\,K$). For Z=0.0002 we built isochrones with the latest OPAL
opacity tables (Rogers, private communication, and Rogers \& Iglesias
1992; Grevesse \& Noels 1993 solar metal ratios) combined with the
molecular opacities of Alexander \& Fergusson 
(1994). The different choices for the opacity tables  is not
relevant because, as we will discuss in Sect.~4, the LFs are
completely unaffected by the adopted opacity coefficients. 
For the equation of
state (EOS) we considered two separate regions: an high-temperature
region (T$>$ 10$^6$ K), where matter can be assumed to be completely
ionized and where we adopted the EOS of Straniero (1988) and a low
temperature region (T $<$ 10$^6$ K) where partial ionization takes
place. In this last region the thermodynamical properties of partially
ionized matter are derived from the Saha equation as described in
Chieffi \& Straniero (1989); the pressure ionization is included
according to the method described by Ratcliff (1987).
The colour transformation of Kurucz (1992) was used to transform from
the theoretical temperatures and luminosities to the UBVRI system.

For a comprehensive discussion about the comparison between
theoretical and observational luminosity functions, we refer to
Ratcliff (1987) and references therein.  Here we just wish to note that,
despite of some evident advantages of LFs such as the fact that they
are almost independent
of the unknown details of model envelope structure, the most
constraining disadvantage of this method is the requirement of a
complete count of stars down to very faint magnitudes ($M_V \ge20$).
This is the main reason why luminosity functions are not
used very frequently. During the last years, the situation has been
improved  with the availabily of CCDs and related software packages;
however, it still is not possible to claim that the main problems have been
solved completely.

 The difficulties in building observational
luminosity functions include: the problem of lack of completeness at
low magnitudes (even if modern techniques are available to make
a quantitative evaluation of the completeness, see e.g. Bolte 1989),
 the proper
normalization between various data sets to build the total luminosity
function of a cluster, the removal of background and foreground
objects, crowding, and possible systematic errors which could occur
during the process of data reduction. It is also important to mention
statistical 
noise: to find all possible features in the subgiant region
one needs bins as narrow as  0\fm2 with  a sufficiently large number
of stars such that 
the stochastic variations become smaller than about 10\% (see
e.g. Chieffi \& Gratton 1986). At present, at least to our knowledge,
there are few cluster data available which fulfill these
requirements. Usually observational data are presented with the
statistical error only, but the real errors could be higher.
In all cases we are using data already prepared for LFs, i.e. we use
the number of stars in brightness bins, where corrections for
completeness had been applied by the observers.

\section{Standard luminosity functions of individual globular clusters}

\subsection{M30}

\begin{figure}
\centerline{\epsfxsize=0.65\hsize \epsfbox{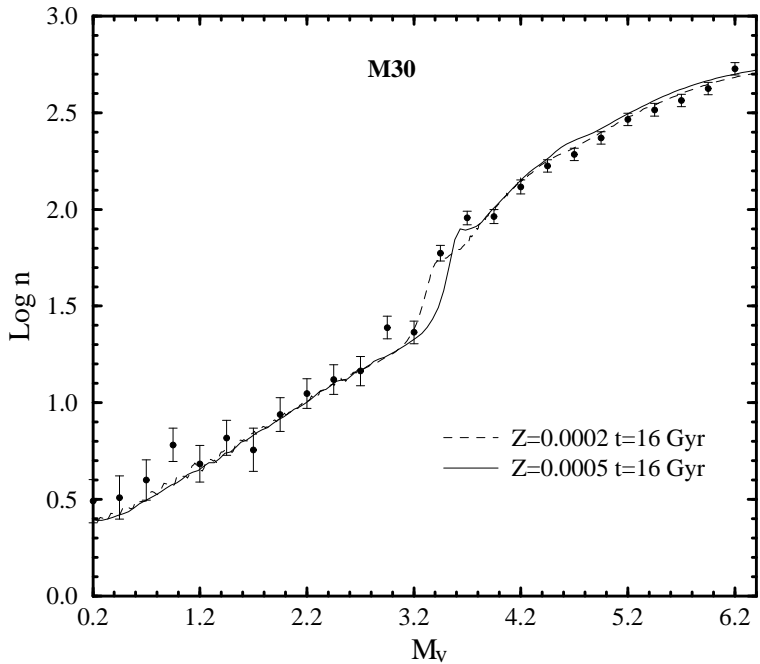}}
\caption[]{The observed luminosity function (LF) for M30 (Bolte 1994;
symbols) and theoretical LFs for $\delta m = 14.65$, an age of 16
Gyr and two metallicities.  
The uncertainty assigned to each bin in the LF is a combination
of Poisson noise and the uncertainty in calculating the correction for
incompleteness as discussed in Bolte (1989)}
\protect\label{m30-lf1}
\end{figure}

We compared the observational LF of M30 taken from Bolte (1994) with
our theoretical standard LF. The composition was $Y=0.23$ and
$Z=0.0005$, where $Z$ is the total metallicity equivalent to that of
the Bergbusch \& VandenBerg (1992) evolutionary tracks with
$[Fe/H]=-2.03$ and $[O/Fe]=+0.7$, which were used by Bolte (1994). 
The distance modulus is $\delta
m=14.65$ and the age is 16 Gyr. For the IMF we chose
$s=2$. These parameters are the same as in the original papers.
The result is shown in Fig.~\ref{m30-lf1} (solid line), where we have
normalized to the RGB. Our LF agrees with the observed values at all
points within $2-3\sigma$, in contrast to the results of Faulkner \&
Swenson (1993). With the same normalization method, IMF slope and
chemical composition, Bolte
(1994) finds a systematic underabundance of observed main sequence
stars. While our LF is only slightly too rich in main sequence stars, 
Bolte finds a more severe lack of stars extending from the turn-off
down the main sequence.
We also show a LF with a lower metallicity of $Z=0.0002$ (dashed
line), which is 
obtained if oxygen- resp. $\alpha$-enhancement is ignored. Actually,
Djorgovski (1993) gives $Z=10^{-4}$ (without $\alpha$-enhancement)
based on results by Zinn \& West (1984) and Armandroff \& Zinn (1988).
Evidently, this fit is satisfying as well.
Our result for M30 already indicates  that the basis for postulating
non-standard physics for low-mass star evolution might not exist.

\subsection{M92, M68, NGC6397}

\begin{figure}
\centerline{\epsfxsize=0.65\hsize \epsfbox{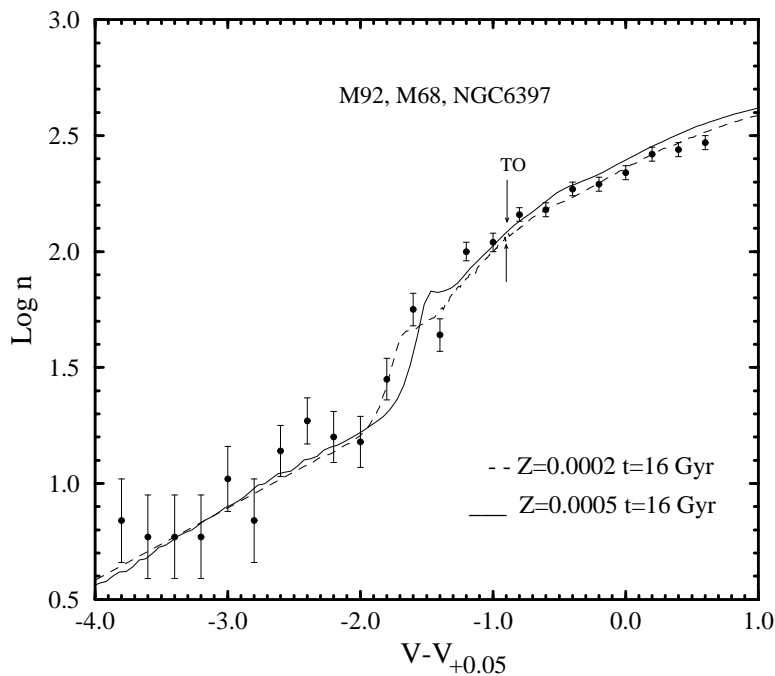}}
\caption[]{The combined observed LFs of M92, M68 and NGC6397 (Stetson
1991) and two LFs for age 16 Gyr and $s=2$.
Errorbars reflect purely statistical errors. As described in Stetson (1991)
the origin of magnitudes has been arbitrarily shifted to the point
on the upper main sequence which is 0\fm05 redder than the bluest colour
at the turn off (indicated by TO)
}
\protect\label{m92-lf1}
\end{figure}

As in Stetson (1991) we created a composed LF of the three globular
clusters M92, M68, NGC6397 (Fig.~\ref{m92-lf1}), which show very similar
CMDs. A necessary condition for this procedure is that the clusters
have nearly identical metallicity, reddening, age and IMF (here
$s=2$). Then, one can 
assume that at a given point (Stetson chose that the one being 0\fm05 redder
than the bluest colour at the turn off) they also have the same
absolute brightness and their CMDs can be superimposed. For an assumed
age (here 16 Gyr) the same procedure applies for the theoretical isochrone.
Although we have followed Stetson (1991), we think that such a
combination of LFs is not advisable.
Again, we compared the observations with our standard LFs for two
metallicities,  
$Z = 0.0005$ and $0.0002$. The higher value, as well as all other
parameters is in agreement with Stetson (1991).
Within $2-3\sigma$ (statistical errors only) theory matches
observations; the fit employing the lower metallicity is
slightly better. As in the case of M30 we cannot detect any
significant discrepancy. 

\subsection{M4 and NGC1851}

\begin{figure}
\centerline{\epsfxsize=0.65\hsize \epsfbox{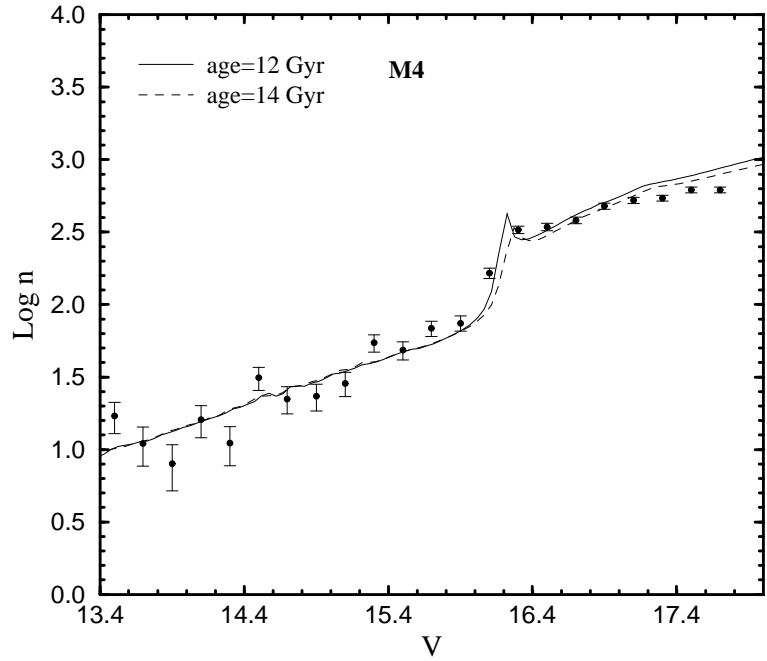}}
\caption[]{The observed LF of M4 (Piotto \& Saviane 1996)
and two theoretical LFs for $\delta m=12.75$, $Z = 10^{-3}$ and age
12 (solid) and 14 (dashed) Gyr, the latter one shifted by
0\fm1. Errors are purely statistical} 
\protect\label{m4-lf1}
\end{figure}

For these and the following clusters the data are from Piotto \&
Saviane (1996) and the distance modulus (here: $\delta m = 12.75$)
from Djorgovski (1993). 
For M4 we found the best agreement
for the parameters $Z = 10^{-3}$ ($9\,10^{-4}$ in Djorgovski 1993), $s =
1$, and $t= 14\, {\rm Gyr}$ (shifted by 0\fm1). This age is in
agreement with Chaboyer \& 
Kim (1995). For comparison we also show the LF for an
age of 12 Gyr, but otherwise the same
parameters. The observational data are 
well reproduced by both theoretical LFs (Fig.~\ref{m4-lf1}).
The 12 Gyr LF fits better to the subgiant region, the 14 Gyr one to the
main sequence. 

\begin{figure}
\centerline{\epsfxsize=0.65\hsize \epsfbox{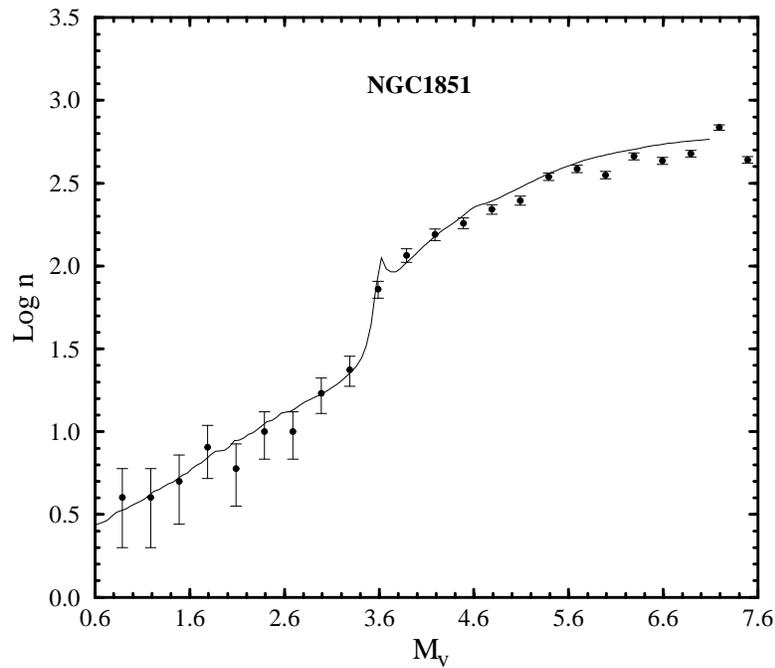}}
\caption[]{The observed LF of NGC1851 (Piotto \& Saviane 1996)
and our theoretical LF for $s=2.35$, $\delta m=15.46$, $Z = 10^{-3}$ and age
14 Gyr. Errors are purely statistical}
\protect\label{n1851-lf1}
\end{figure}

NGC1851 (Fig.~\ref{n1851-lf1})
can be fitted by our theoretical LF with the 
parameters $\delta m=15.46$, $Z = 10^{-3}$ (Djorgovski 1993)
and age 14 Gyr (Chaboyer \& Kim 1995 give $12.6\pm 1.6$ Gyr). 
The data  are given as
absolute visual magnitudes. As in the preceeding cases, there are too
few main sequence stars observed (but the counts are still within
$2-3\sigma$), which we suggest to result from uncorrected incompleteness.

\subsection{NGC6352}

\begin{figure}
\centerline{\epsfxsize=0.65\hsize \epsfbox{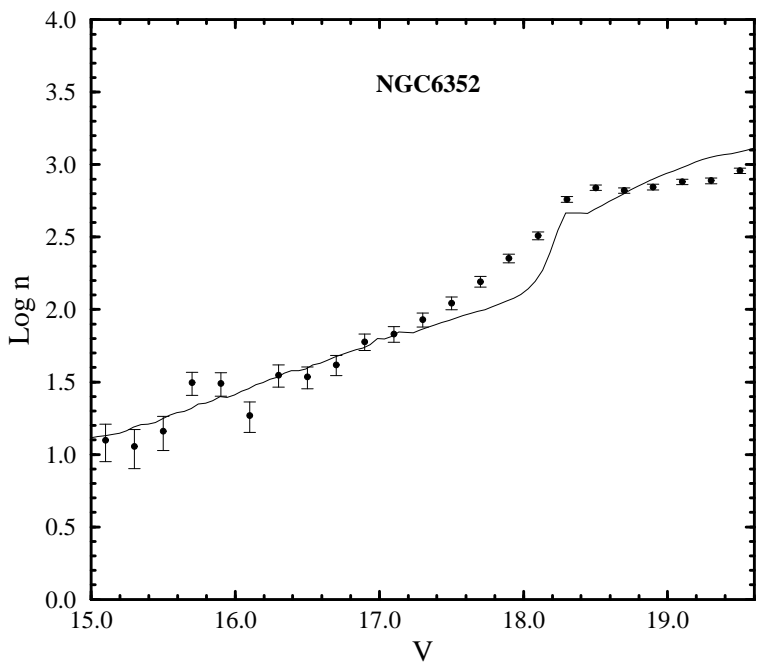}}
\caption[]{The observed LF of NGC6352 (Piotto \& Saviane 1996)
and our ``best'' theoretical LF for $\delta m=14.52$, $s=1$, $Z =
10^{-3}$ and age 18 Gyr (statistical errors only)}
\protect\label{n6352-lf1}
\end{figure}

For this cluster  we could not
obtain a satisfying fit. Even our best case, shown in
Fig.~\ref{n6352-lf1}, is comparably poor for both the subgiant and main
sequence region, where we have an over- resp. underdensity of observed
stars. The slope of the subgiant to giant branch is smaller than
theoretically predicted. According to the dependencies investigated by
Ratcliff (1987), such a shape of the LF could indicate a very
low metallicity. However, the metallicity used is
$Z = 0.001$, which is already lower than Djorgovski's (1993)
estimate of 0.006. Using this value results in a definitely worse LF
fit. We have tested very low metallicities, but the fit does not
improve significantly; the same being true for age variations.
We conclude that this cluster needs substantially better
observations and an accurate determination of its metallicity before one
can compare it with theoretical LFs.

\subsection{M80}

\begin{figure}
\centerline{\epsfxsize=0.65\hsize \epsfbox{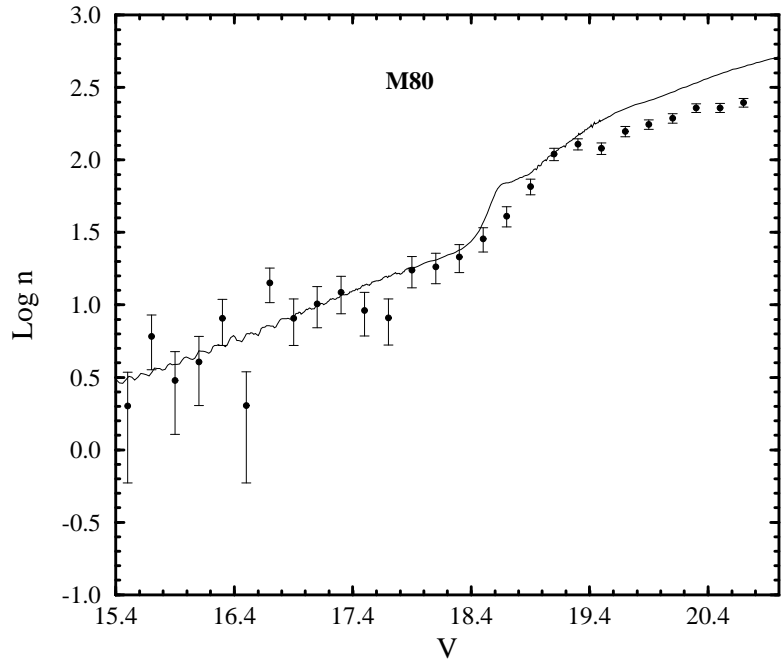}}
\caption[]{The observed LF of M80 (Piotto \& Saviane 1996)
and a theoretical LF for $\delta m=15.24$, $Z = 2\,10^{-4}$ and age
16 Gyr. Notice the discrepancy at $V\approx 18.8$. Errors are
statistical ones only}
\protect\label{m80-lf1a}
\end{figure}

\begin{figure}
\centerline{\epsfxsize=0.65\hsize \epsfbox{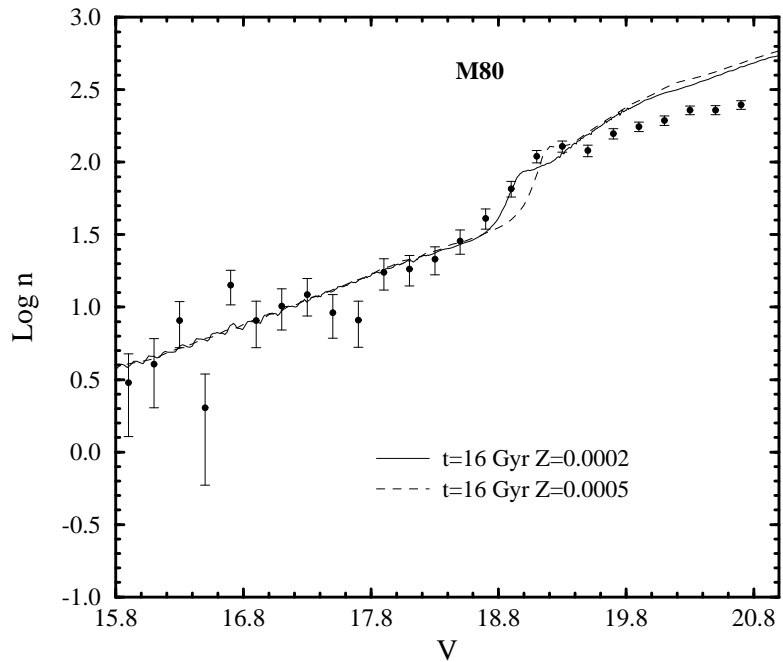}}
\caption[]{The same as Fig.~\ref{m80-lf1a} (solid line), but with a
distance modulus of $\delta m=15.57$. Also shown is a LF with higher
metallicity, which could arise, if the cluster is  significantly
enhanced in $\alpha$-elements}
\protect\label{m80-lf1b}
\end{figure}

Our last cluster is M80. For the
choice of parameters (age 16 Gyr, $Y= 0.24$,  $s=2$, $Z= 2\,10^{-4}$;
Djorgovski 1993 gave  $Z= 4\,10^{-4}$) and the distance modulus of
Djorgovski (1993) of $\delta m = 15.24$ we obtain a 
bad fit with a severe overdensity of subgiants and main-sequence stars
(Fig.~\ref{m80-lf1a}). However, the structure of the observed LF
indicates that the bump at $V \approx 19.2$ is the true location of
the subgiants. We therefore have changed the distance modulus to
$\delta m = 15.57$ (Scotti 1995).
The resulting LF is shown in Fig.~\ref{m80-lf1b}
for the same and a higher metallicity. Evidently, it fits much better,
although there still is an overdensity of main-sequence stars. This
example illustrates how a ``discrepancy'' may arise from a wrong
distance modulus.

\section{Luminosity functions with varying assumptions}

Up to now our theoretical luminosity functions have been compared with
the observed ones for several globular clusters, discussing the
possible sources of uncertainty related to the observed LFs.  However,
to get an idea of how reliable such  a comparison is, one must also
analyse the possible uncertainties in the theoretical LFs. For this
purpose we tested the influence of variations in the
main physical inputs quantities (opacities, chemical composition, age, equation
of state, etc.) on the luminosity functions, but also that of an hypothesized
isothermal core as suggested by
Faulkner \& Stetson (1993) and Bolte (1994).  Some of these
parameters have already been discussed in
several papers (see e.g. Chieffi 1986, Ratcliff 1987 and references
therein); in these cases we only summarize the results, since the
basic picture has not changed by the recent improvements in opacities
and the EOS. A compilation of these tests can be found in Leone (1995).

\subsection{Input physics variations}

To test the influence of the different calculations of
radiative {\sl opacity} coefficients on LFs we compared LFs constructed
with all the (four) possible combinations of choosing
the Los Alamos opacity tables (Huebner et al. 1977) or the updated
OPAL opacities (Rogers \& Iglesias 1992; Rogers 1994, private 
communication) for the interior opacities (for temperatures above 10$^4$
K) and of choosing Cox \& Tabor (1976) or the new Alexander \&
Fergusson (1994) molecular opacity tables
for the cool external regions. It is well known from e.g. solar
models and homology considerations that opacity influences the
stellar luminosity of main sequence models and thereby their
lifetimes; 
however, this results only in a global shift of the LF; a change 
in the luminosity function's shape is not to be expected, since the
shell luminosity depends only weakly on the opacity. Indeed, we did
not find any appreciable differences in the various LFs. 

We also compared LFs constructed by using the {\sl EOS} of Straniero (1988)
with those where the updated Livermore EOS
(Rogers 1994; Rogers, Swenson \& Iglesias 1996) had been used. The
latter one  is based on an approach which
avoids an ad hoc treatment of the pressure ionization and includes
also subtle quantum effects in the corrections for Coulomb forces. Again, 
the influence is negligible. Note, however, that the new OPAL EOS leads
to a reduction of GC ages (Chaboyer \& Kim 1995;
Degl'Innocenti, Salaris \& Weiss 1996) and therefore also influences LFs
indirectly.

One physical parameter which may influence the
stellar evolution is the efficiency of the {\sl nuclear reaction
rates}. The rate of the proton-proton reaction is too low to be
directly measured in the laboratory and it can be determined only
theoretically, while for most of the other important hydrogen burning
reactions the adopted astrophysical factors
are extrapolations of experimental data
taken at energies higher than those relevant for stellar
interiors (see e.g. Rolfs \& Rodney 1988).
The uncertainties usually claimed for the rate of the reactions
which drive the H burning are below 5\% for
the reactions of the pp chain (somewhat higher for $^7{\rm
Be}+{\rm p}$) and
smaller than about 15\% for the reactions of the CNO bi-cycle. 
A variation in the efficiency of these reactions leads to a minor 
change in the CNO-burning temperature but does not influence the
stellar luminosity or evolutionary timescales as can be inferred from
homolgy considerations of shell burning stars.
We constructed LFs where the
reaction rates for the p+p, $^{3}$He+$^{3}$He and
$^{14}$N+p, resp., were changed by a constant factor beyond
the experimental errors. These LFs remained almost unaffected. Only in
the case of the $^{14}$N+p reaction modified by a huge
(hypothetical) factor of 5, we
found a shift in the luminosity of the LF, which, however, is still
within the range of the globular
distance modulus uncertainty. Since a global change in the evolutionary
speed cannot be discriminated from a distance variation, we
had chosen this reaction for this test, because the CNO-cycle
becomes important only when the hydrogen shell is developing, such that
the influence is larger after the turn-off than on the main sequence.
Nevertheless, the change is too small to be important.

\subsection{Influence of chemical composition}

Ratcliff (1987) already pointed out that the dependence of
the LFs on the {\sl helium content} is not high and cannot be
separated cleanly from that of the unknown IMF.
Here we checked that a change in $Y$ of $\pm$ 0.01 w.r.t.\ 
a central value of $Y=0.23$ (which reflects the uncertainty in the
helium content of galactic globulars) has no detectable influence at
all on the LFs.

It is well known (see e.g. Paczynski 1984,
Ratcliff 1987) that the {\sl metallicity content} strongly affects either
the position or the slope of the subgiant branch of the luminosity 
function. We convinced ourselves that a change in $[Fe/H]$ of $\pm$ 0.2 dex,
which is much larger than the average spread in metallicity
of any given cluster (see, e.g. Suntzeff 1993 and references therein), 
gives an effect which can be corrected by shifting
the LF by less than 0\fm1, which in turn is less than the 
uncertainty in the distance modulus. 

All our LFs have a scaled solar {\sl metal composition},
while a growing amount of observational data 
shows that low metallicity globular clusters are
very probably enhanced in all of the $\alpha$-elements
(see e.g. Lambert 1989).
Salaris  et al. (1993) demonstrated that the isochrones
for $\alpha$-enhanced composition are the same as scaled solar
LFs with the corresponding global metallicity. In addition, metal ratios
affect evolutionary speed -- if at all -- only via the CNO-cycle
efficiency, similar to the correspondig reaction rates (see
above). Therefore, no influence on LFs is expected and solar-scaled
metal mixtures can be used savely.

We finally add that Proffit \& VandenBerg (1991) pointed out that
differences in the shapes of the luminosity functions
between canonical models and those which include {\sl helium
diffusion} are too small to be observationally detectable.

\subsection{Other parameters: age, IMF and mixing length}

All these parameters have been extensively dicussed
by Ratcliff (1987) and Paczynski (1984). 
They noted that the assumed initial mass function affects only the 
main sequence part of the LFs, while the LFs are almost independent
of the adopted treatment of convection, in particular of the choice of
the mixing length parameter (because convection only influences
surface temperature as do low-temperature opacities).

On the contrary, the position of the subgiant region of the LF is 
strongly dependent on age; it could therefore be used as an
independent way of determining GC ages or to check that ages
determined by use of CMDs result in consistent LFs. In all our cases
we indeed have used ages consistent with
determined values. Additionally, we checked that a variation 
in age of $\pm$ 2 Gyr, that is within the claimed
uncertainty of determined globular cluster ages, can again be
compensated by a shift in luminosity of about 0\fm1, which is well within
the distance uncertainty. 

We can thus conclude that the shape of the LFs is well defined
theoretically, being little sensitive to the usual uncertainties
in the physical inputs  used in the theoretical models of globular cluster
stars. In particular we emphasize that the luminosity functions are not
affected at all by the uncertainties in mixing length parameter and
complicated low temperature opacities which constitute one of the main
problem in the calculation of theoretical CMDs.

\subsection{Additional energy transport in core}

Within the standard scenario we found, as already discussed in
Sect.~3, a good  fit for most clusters we examined
except for NGC6352, whose LF shows a very different behaviour with
respect to the theoretical one both in the subgiant and main sequence
region. For all other clusters, we could not find any evidence for a
significant 
discrepancy; in fact, we would claim that the theoretical LFs fit the
observations very well. Since
Faulkner \& Swenson (1993) and Bolte (1994) found that their fits improved
for stellar models with an isothermal core in the innermost 10\% in
mass, we also investigated this non-standard change in the theoretical
models. 
To this end, we closely followed the procedure by Faulkner \& Swenson
(1993) and evolved stellar models under the assumption of a strongly
enhanced energy transport in the central 10\% of their mass. We
achieved this by reducing the radiative opacity by a factor of
$10^{-4}$ artificially. Otherwise, the assumptions, parameters and the
procedure 
for creating the LFs is the same as in Sect.~3. 

\begin{figure}
\centerline{\epsfxsize=0.65\hsize \epsfbox{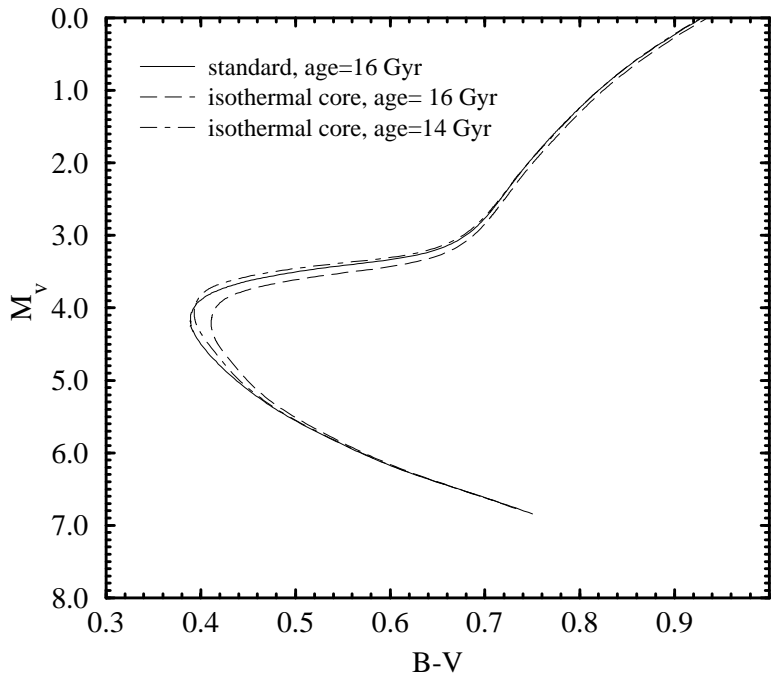}}
\caption[]{Comparison between a standard isochrone of 16 Gyr
and two isochrones (of the indicated ages) obtained from models
with an isothermal core in the
innermost 10\% in mass (Z=0.0002, Y=0.23)}
\protect\label{iso-iso}
\end{figure}

Before we turn to the LFs, we wish to address an additional point made
by Faulkner \& Swenson (1993), who stated that in the
``isothermal core'' case an age reduction of the globular clusters by
about 20\% would result. This conclusion was based on the comparison
of the turn-off (TO) age of two models of $M= 0.80 M_\odot$
(standard)  and of $M= 0.8185 M_\odot$ (with isothermal core) 
which have the same TO temperature.
We decided to calculate both standard and isothermal
core isochrones (Fig.~\ref{iso-iso}). If the age is the same for both
(16 Gyr), they  have 
the same TO visual magnitude and a sightly different (B-V) by about
0\fm02. 
On the other hand, an isothermal core isochrone of 14 Gyr has the same TO
colour as the standard one, but a different visual magnitude ($\Delta M_V
\approx$ 0\fm1).

There are two main
procedures to determine the age of galactic globulars
(see e.g. Demarque et al.\ 1991; Salaris et al.\ 1993 and references
therein):
the $\Delta$V(HB-TO) method, which uses the $M_V$ difference between
the TO and the HB at the level of the RR Lyrae stars, and the 
$\Delta$(B-V) method, which uses the (B-V) colour difference 
between the TO and the base of the RGB. Both these procedures are
claimed to have
an internal error of about $\pm$ 2.5 Gyr (see e.g. Salaris et
al. 1993). Depending on what method one prefers, isochrones with
identical TO luminosity resp.\ effective temperature have to be
compared, and therefore an age reduction will or will not be found for
the isothermal case.
For the $\Delta$V(HB-TO) method, we comment that
modelling evolutionary tracks with a 10\% isothermal core until the He
ignition
does not influence at all the following horizontal branch evolution,
thus the horizontal branch luminosity is supposed to be the same as
the standard one (Dearborn et al 1990).

It is evident from Fig.~\ref{iso-iso} that both the 14 
and the 16 Gyr isothermal isochrones fit the standard one
within the estimated uncertainties. However, if  forced to choose, 
we would prefer the one of higher age, i.e. the one with the same TO
luminosity. The reason is that
the determination of the TO colour is a difficult procedure
affected by our poor knowledge of stellar envelope models (e.g. 
mixing length formulation of convective energy transport, $T_{\rm{eff}}$-(V-B)
relation, low temperature opacities). A GC age reduction can only be
claimed if the TO temperature is predefined. Otherwise, the isothermal
core isochrones give the same age as the standard ones.

\begin{figure}
\centerline{\epsfxsize=0.65\hsize \epsfbox{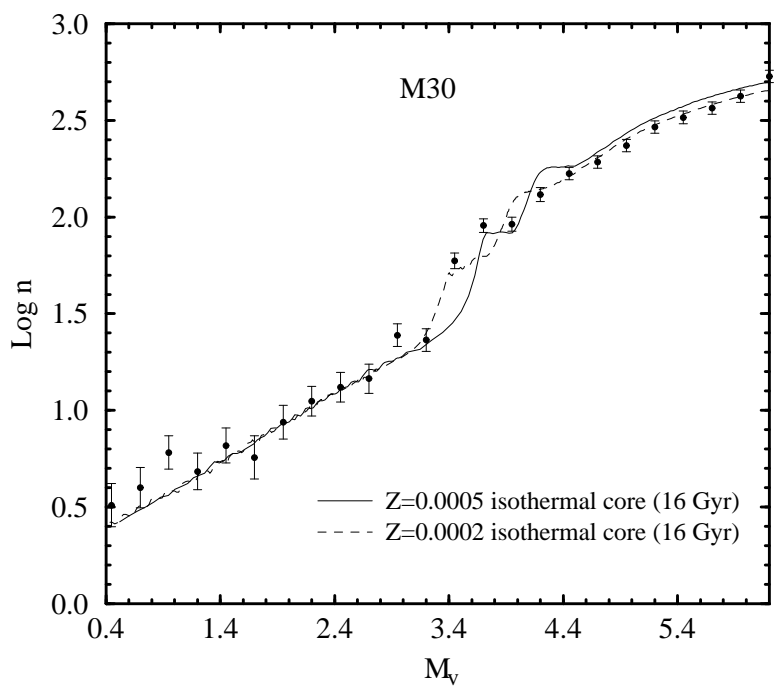}}
\caption[]{Comparison between the observed luminosity function of M30 (as
in Fig.~\ref{m30-lf1}) and our  ``isothermal'' LFs for
$Z=0.0005$ (solid line) resp.\ 0.0002 (dashed). Note the additional bump
at $M_V \approx 4.2$. All other parameters are as in Fig.~1}
\protect\label{m30-lf2}
\end{figure}

\begin{figure}
\centerline{\epsfxsize=0.65\hsize \epsfbox{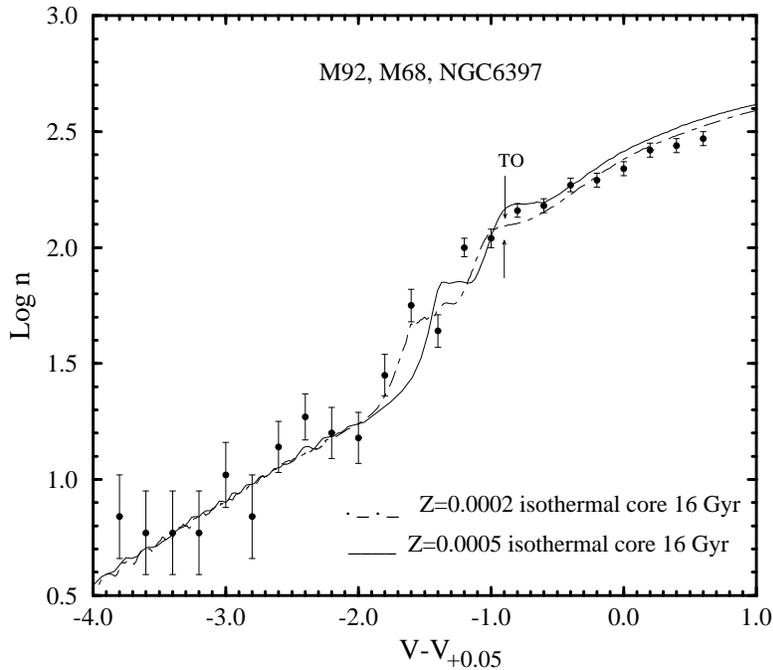}}
\caption[]{Comparison between the combined observed luminosity
function of M92, M68 and NGC6397 (as
in Fig.~\ref{m92-lf1}) and our ``isothermal'' LFs for $Z=0.0005$ (solid) and 0.0002
(dashed)} 
\protect\label{m92-lf2}
\end{figure}

We now compare our ``isothermal'' LFs with the observed
ones; for each cluster the chemical composition, the IMF and the
distance modulus are as in Sect.~3. 
Fig.~\ref{m30-lf2} shows our comparison for M30 for the same age and both
metallicities. 
In disagreement with Faulkner \& Swenson (1993)
and Bolte (1994), even these ``best'' fits do not appear to be an
improvement over
the standard ones (Fig.~\ref{m30-lf1}). In fact, we think they are worse
because of the additional bump shortly after the TO.

Figure \ref{m92-lf2} shows the comparison between observational data and
isothermal LFs for the composed LF of M92, M68, NGC6397.
Again, age and metallicities are chosen as in the standard case.  Neither
in this case do the fits improve over the standard one.

\begin{figure}
\centerline{\epsfxsize=0.65\hsize \epsfbox{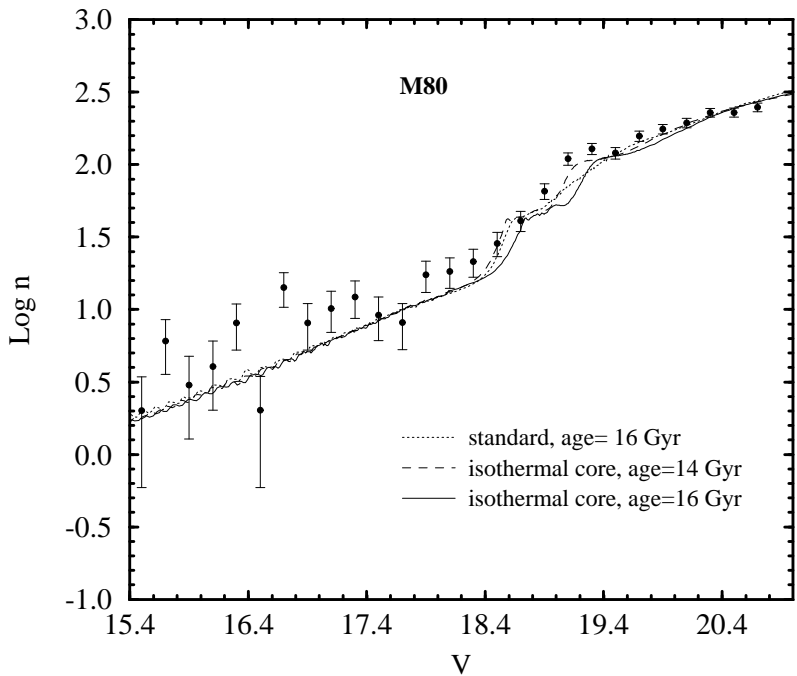}}
\caption[]{Comparison between the observed luminosity
function of M80  and our theoretical ones, when normalized to the main
sequence. The distance modulus is $\delta m = 15.24$. Shown are the
standard case (dotted) and two ``isothermal'' LFs 
of 16 (solid) and 14 Gyr (dashed); other parameters are as in
Fig.~\ref{m80-lf1a} ($Z=0.0002$)}
\protect\label{m80-lf2}
\end{figure}

For all other clusters, the isothermal LFs are worse than the standard
ones as well. Neither can NGC6352 be improved by the assumed energy transport in
the stellar interior. We therefore are again in disagreement with
Faulkner \& Swenson (1993).
However, we finally wish to illustrate in the case of M80 that the choice
of normalization 
is very important for the conclusions about fit quality.
In Fig.~\ref{m80-lf2} we show fits obtained by normalizing to the main
sequence. The dotted line is the standard case with parameters as in
Fig.~\ref{m80-lf1a}, i.e. with the smaller distance modulus. Around the
TO, there appears to be a deficit in the theoretical LF. The solid line
is a fit with an isothermal core of the same age, which already improves
the fit. At last, a fit with an isothermal core and an age of only 14 Gyr
(dashed) produces a very good fit, and one would conclude from this that
both an  additional energy transport in the core (by WIMPs) and a younger
age are indicated or even necessary. However, we have shown in Sect.~3
that a sufficiently good fit is also obtained by a standard LF and a
larger distance modulus. 
 
\section{Discussion and conclusions}

\begin{figure}
\centerline{\epsfxsize=0.65\hsize \epsfbox{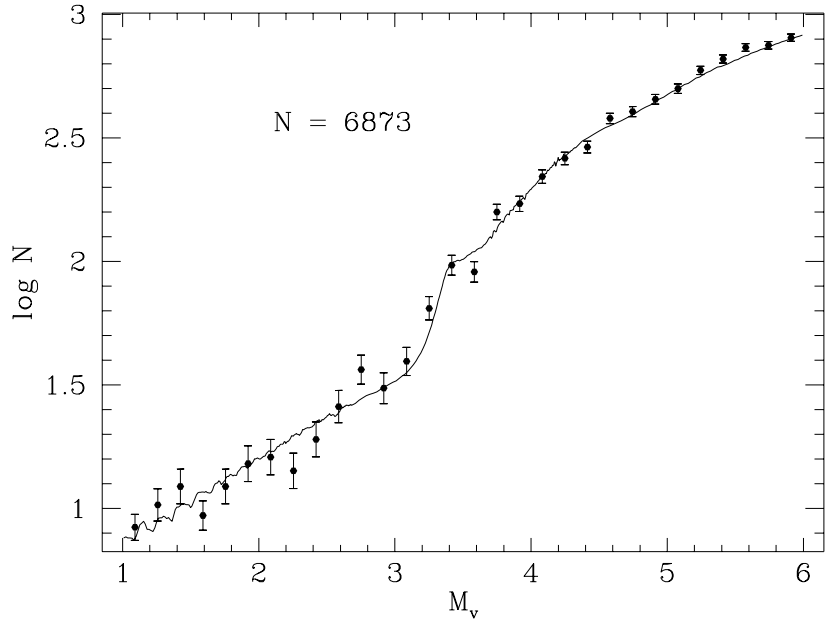}}
\caption[]{Theoretical and synthetic luminosity functions for standard
parameters (16 Gyr; $s=2$, $Z=2\, 10^{-4}$). 26000 random points in the
N-$M_V$ plane were chosen, out of which 6873 fell below the theoretical
probability function and were accepted. The number of brightness bins is
30. $1\sigma$ errors are indicated
}
\protect\label{syn-lf}
\end{figure}

The prime intention of this paper has been to independently investigate
whether there indeed exists a systematic and significant discrepancy
between observed and theoretical luminosity functions of Globular
Clusters as claimed by Faulkner \& Stetson (1993) and Bolte (1994). We
have demonstrated in Sect.~3 that for the clusters used in their papers
and for additional data by Piotto \& Saviane (1996) the agreement between
our standard luminosity functions and the observations is very good for
reasonable assumptions about cluster composition, distance and
age. Actually, we have used values from the literature for these
parameters, but exploited the range of uncertainties to find the best
fit. All data points can be fitted within $3\sigma$ of the statistical
errors, except for the main sequences. 
To illustrate the influence of pure number statistics we have performed
the following test: we took one of our standard LFs (16 Gyr; $s=2$,
$Z=2\, 10^{-4}$) and constructed a synthesized LF by the rejection method
to construct random deviates. Fig.~\ref{syn-lf} shows the comparison
between the synthetic and the theoretical LF for a total number of stars
and a number of brightness bins comparable to the cases discussed in this
paper. The similarity, for example with Fig.~\ref{m30-lf1}, demonstrates
that our fits are perfect within the statistical errors, and that
deviations like those at $M_V = 3.6$ or $3.8$ are to expected. In fact,
some of our cases show a better agreement than would be consistent with
the statistics. In these cases, the fit parameters might be
overdetermined. Of course, the systematic deviations on the main
sequences are a clear indication of non-statistical errors.
Since all clusters by  Piotto \& Saviane (1996) are deficient with
respect to the main-sequence parts of the theoretical LFs, while
those used by Faulkner \& Stetson (1993) are not, we ascribe this to an
underestimate in the completeness-correction applied by the observers.

Alternatively, one could speculate that the IMF for these clusters is
flatter than assumed. This uncertainty -- and the
additional one of the helium content -- prevented us from using the main
sequence for the normalization of the LFs, in spite of the much smaller
statistical errors. Instead, we used the RGB for normalization, whose
LF-slope is an extremely stable quantity independent of all
input variables. In the case of M80 we demonstrated that the use of the
main sequence would lead to a ``discrepancy'' in the standard LF and its
resolution by an isothermal core, just as was the case in the papers
triggering the present work.

The main sequence being inadequate for a detailed comparison and the red
giant branch being robust against model changes, the subgiant branch is
left to reflect model differences, parameter influences and potential
problems. However, the present observations do not resolve the LF in this
region sufficiently well and with adequate accuracy. We propose that in
the future observations aiming at obtaining the LFs of GCs should
concentrate exactly on this region.

We found one cluster (NGC6352), which withstands all attempts to fit a
theoretical LF. The shape of the observed one is in fact very strange and
looks like that of a very low-metallicity system in terms of the missing
subgiant break, and like one of high helium content with respect to the
slope at the TO (Ratcliff 1987). However, within the parameter range
investigated by us, we could not obtain any good fit. We rather suspect
that the subgiant break was not resolved properly, and this should be
checked by further observations. 

Of all quantities we have tested for their influence on the LFs, we found
that metallicity and distance (see the example of M80;
Figs.~\ref{m80-lf1a} and \ref{m80-lf1b}) have the largest influence. Age
variations of up to 2 Gyr can be compensated by a luminosity shift
smaller than the quoted distance uncertainty (because old clusters change
their TO-luminosity hardly with age). Discrepancies in the LF should
therefore first raise the question of correct distance or metallicity
determinations. In this context information about $\alpha$-element
enrichment is important, too.

In Sect.~4 we repeated the LF-calculations, but with a very efficient
energy transport in the innermost 10\% of our models (following again
Faulkner \& Swenson 1993 and Bolte 1995). Our results show that the fits
are not improved at all, but rather get worse.
Since we found no evidence for a LF-discrepancy, it is not necessary to
discuss properties of hypothetical WIMPs for additional energy transport
in the core of main-sequence stars.

To summarize, we have shown that the agreement between observed and
theoretical luminosity functions appears to be very good, with problems
only arising on the lower main sequence (corrections for completeness
might have to be improved) and for the exact shape of the subgiant bump
and break (resolution in brightness), which have the potential to yield
information about metallicity, age and distance of the cluster.

\begin{acknowledgements}
We are grateful to D.~Alexander and F.~Rogers for making available
their opacity tables to us. 
Thanks also go to G. Piotto and I. Saviane for providing us
with their observational data before  publication and 
to A. Chieffi and O. Straniero for making available to us
their calculated isochrones.
We appreciate instructive discussions with
G.~Raffelt and V.~Castellani and the careful reading of the manuscript
by H.~Ritter, who also directed our attention to homology relations
for shell burning stars.
Part of the work of A.W.\ and S.D'I. was supported 
by the ``Sonderforschungsbereich 375 f\"ur
Astro-Teilchenphysik der Deutsche Forschungsgemeinschaft''.
\end{acknowledgements}

{}

\end{document}